\documentclass[12pt,preprint]{emulateapj}

\usepackage{graphicx,amsmath,natbib}

\slugcomment{Accepted to ApJ, Oct. 12$^{th}$, 2010}

\shorttitle{An improved orbit for the $\alpha$ Oph System}
\shortauthors{Hinkley et al.}

\begin{document}

\title{Establishing $\alpha$ Oph as a Prototype Rotator: Improved Astrometric Orbit}

\author{Sasha Hinkley\altaffilmark{1,17}}
\author{John D. Monnier\altaffilmark{2}}
\author{Ben R. Oppenheimer\altaffilmark{3}} 
\author{Lewis C Roberts Jr.\altaffilmark{4}}
\author{Michael Ireland\altaffilmark{7}}
\author{Neil Zimmerman\altaffilmark{5,3}}
\author{Douglas Brenner\altaffilmark{3}}
\author{Ian R. Parry\altaffilmark{6}}
\author{Frantz Martinache\altaffilmark{8}}
\author{Olivier Lai\altaffilmark{9}}
\author{R\'emi Soummer\altaffilmark{10}}
\author{Anand Sivaramakrishnan\altaffilmark{3,11,12}}
\author{Charles Beichman\altaffilmark{14}}
\author{Lynne Hillenbrand\altaffilmark{1}}
\author{Ming Zhao\altaffilmark{4}}
\author{James P. Lloyd\altaffilmark{15}}
\author{David Bernat\altaffilmark{15}}
\author{Gautam Vasisht\altaffilmark{4}}
\author{Justin R. Crepp\altaffilmark{1}}
\author{Laurent Pueyo\altaffilmark{4}}
\author{Michael Shao\altaffilmark{4}}
\author{Marshall D. Perrin\altaffilmark{16,18}}
\author{David L. King\altaffilmark{6}}
\author{Antonin Bouchez\altaffilmark{13}}
\author{Jennifer E. Roberts\altaffilmark{4}}
\author{Richard Dekany\altaffilmark{13}}
\author{Rick Burruss\altaffilmark{4}}

\altaffiltext{1}{Department of Astronomy, California Institute of Technology, 1200 E. California Blvd, MC 249-17, Pasadena, CA 91125}
\altaffiltext{2}{University of Michigan, Astronomy Department, 941 Dennison Bldg, Ann Arbor, MI 48109-1090}
\altaffiltext{3}{Astrophysics Department, American Museum of Natural History, Central Park West at 79th Street, New York, NY 10024}
\altaffiltext{4}{Jet Propulsion Laboratory, California Institute of Technology, 4800 Oak Grove Dr., Pasadena CA 91109}
\altaffiltext{5}{Department of Astronomy, Columbia University, 550 West 120th Street, New York, NY  10027}
\altaffiltext{6} {Institute of Astronomy, University of Cambridge, Madingley Road, Cambridge CB3 0HA, UK}
\altaffiltext{7}{School of Physics, University of Sydney, NSW 2006, Australia}
\altaffiltext{8}{National Astronomical Observatory of Japan, Subaru Telescope, Hilo, HI, 96720.}
\altaffiltext{9}{CFHT Corp., 65-1238 Mamalahoa Hwy., Kamuela, HI 96743}
\altaffiltext{10}{STScI, 3700 San Martin Drive, Baltimore, MD 21218}
\altaffiltext{11}{Stony Brook University}
\altaffiltext{12}{NSF Center for Adaptive Optics.}
\altaffiltext{13}{Caltech Optical Observatories, California Institute of Technology, Pasadena, CA 91125}
\altaffiltext{14}{NASA Exoplanet Science Institute, California Institute of Technology, Pasadena, CA 91125}
\altaffiltext{15}{Department of Astronomy, Cornell Univ., Ithaca, NY 14853}
\altaffiltext{16}{UCLA Department of Astronomy}
\altaffiltext{17}{Sagan Fellow}
\altaffiltext{18}{NSF Postdoctoral Fellow}


\begin{abstract}
The nearby star $\alpha$ Oph (Ras Alhague) is a rapidly rotating A5IV star spinning at  $\sim$89\% of its breakup velocity. This system has been imaged extensively by interferometric techniques, giving a precise geometric model of the star's oblateness and the resulting temperature variation on the stellar surface.  
Fortuitously, $\alpha$ Oph has a previously known stellar companion, and characterization of the orbit provides an independent, dynamically-based check of both the host star and the companion mass.
Such measurements are crucial to constrain models of such rapidly rotating stars.  In this study, we combine eight years of Adaptive Optics imaging data from the Palomar, AEOS, and CFHT telescopes to derive an improved, astrometric characterization of the companion orbit. We also use photometry from these observations to derive a model-based estimate of the companion mass. A fit was performed on the photocenter motion of this system to extract a component mass ratio.  We find masses of $2.40^{+0.23}_{-0.37}$ M$_\odot$ and $0.85^{+0.06}_{-0.04}$ M$_\odot$ for $\alpha$ Oph A and $\alpha$ Oph B, respectively. Previous orbital studies of this system found a mass too high for this system, inconsistent with stellar evolutionary calculations.  Our measurements of the host star mass are more consistent with these evolutionary calculations, but with slightly higher uncertainties.  In addition to the dynamically-derived masses, we use $IJHK$ photometry to derive a model-based mass for $\alpha$ Oph B, of 0.77$\pm$0.05  M$_\odot$ marginally consistent with the dynamical masses derived from our orbit. 
Our model fits predict a periastron passage on 2012  April 19, with the two components having a ~50 mas separation from March to May 2012. A modest amount of interferometric and radial velocity data during this period could provide a mass determination of this star at the few percent level.

\end{abstract}




\section{Introduction}
Although the binary properties of solar-type stars \citep{dm91,mh09} and lower mass late-type stars \citep{fm92,rg97}, are becoming increasingly well understood, studies of the {\it higher} end of the mass spectrum are crucial for a broader understanding of stellar multiplicity.  Indeed, recent simulations suggest that A stars may possess a greater abundance of both stellar and planetary mass companions \citep[][J. Crepp, private communication]{kmy10}.  However, the multiplicity characteristics of these more massive stars has only recently started to be surveyed \citep{st02,kbz05}, and more extensive surveys are underway to place limits on the nature of massive-star multiplicity.  In recent years, high contrast imaging techniques \citep{oh09,am09} such as adaptive optics (hereafter ``AO'') have matured significantly, allowing these studies to proceed.  The nearby (14.68 pc) A5IV star $\alpha$ Oph (Ras Alhague) has a well known companion \citep{w46,lw66,g05} with a 8.62-year period,  well established over several decades of monitoring and first resolved by \citet{m83}.  But a fuller characterization of the companion has not been possible, and, aside from the period, the orbital parameters have been loosely constrained \citep{klm89,ah92,g05}.

\begin{deluxetable*}{llrlrlcclc}
\tabletypesize{\scriptsize}
\tablecaption{Table of Astrometry}
\tablewidth{0pt}
\tablehead{\colhead{epoch } & \colhead{MJD} & \colhead{r (mas)} & \colhead{$\sigma_r$} & \colhead{PA} & \colhead{$\sigma_{PA}$} &  \colhead{$x,y$ (mas)} & \colhead{Band} & \colhead{Telescope (Instrument)} & \colhead{Ref\footnotemark[1]}}
\startdata                                   
1   & 2445157.9  & ...             & ...                 &   ...         &  ...                             & -380Y$\pm$50    & $K,L$                                       &  3.8m KPNO Mayall            &   1  \\ 
2   & 2445244.8  & ...             & ...                 &   ...         &  ...                              & -650X$\pm$50   & $K,L$                                       &  3.8m KPNO Mayall             &   1 \\
3   & 2445424.2  & ...             & ...                 &    ...        &  ...                              & -340Y$\pm$50   & $K,L$                                        &  3.8m KPNO Mayal               &  1 \\
4   & 2445804.0  & ...             &  ...                &    ...        &  ...                              & -430Y$\pm$50   & $K,L$                                        &  3.8m KPNO Mayall             &   1 \\
5   & 2447045.8  & ...             &  ...                &   ...         &  ...                              & -400X$\pm$50   & $K,L$                                          &  3.8m KPNO Mayall             &  1 \\
6   &  2451304.5 &  770        &  $\pm$ 40  & 243.7    & $\pm$ 3.0$^\circ$  &    ...                          &  2.15$\mu$m                         &  3.6m La Silla (SHARP II)  & 2 \\     
7   &  2452473.8 &  470        & $\pm$ 20   & 233.2    &  $\pm$ 2.4$^\circ$ &    ...                          &  $I$                                          &  AEOS (VisIm)                     & 3 \\      
8   &  2452760.0 &  $<$275 &$\pm$  3     & n/a         &                                   &    ...                          &  $I$                                          &  AEOS (VisIm)                      & 3,4 \\  
9   &  2453168.5 &  303        & $\pm$ 33   & 253.0    &  $\pm$ 6.3$^\circ$ &    ...                          &  $H$                                        &  AEOS (Lyot Project)          & 5  \\    
10 &  2454253.9 &  776.5    & $\pm$ 2.1  & 244.6    &  $\pm$ 0.4$^\circ$  &   ...                           &  $J,H,K$                                 &  Palomar (PHARO)             & 6 \\      
11 &  2454263.5 &  765        & $\pm$ 20  & 243.7    & $\pm$ 1.5$^\circ$   &   ...                           &  $H$                                        &  AEOS (Lyot Project)         & 5 \\       
12 &  2454637.5 &  787.8    & $\pm$ 2.8  & 240.6    &  $\pm$ 0.4$^\circ$  &   ...                           &  $K$, Br$_\gamma$             & Palomar (PHARO)            & 6 \\        
13 &  2454657.5 &  790       & $\pm$ 20   & 239.5    & $\pm$ 1.4$^\circ$   &   ...                           &  $J$,$H$                                 &  Palomar (Project 1640)   & 7 \\        
14 & 2454963.8 &  756.6     & $\pm$ 7.5  & 239.3   & $\pm$ 1.2$^\circ$    &   ...                          &  Pa$_\beta$ (1.28$\mu$m) &  CFHT (PUEO)                   &  8\\        
15 &  2455002.5 &  748.5  & $\pm$ 2.7  & 238.3     & $\pm$ 0.4$^\circ$    &   ...                          &  $K$                                         &  Palomar (PHARO)           &  6 \\        
\enddata
\footnotetext[1]{References: 1) \citet{m83} and \citet{klm89} provide one-dimensional speckle measurements 2) \citet{bmm01}, 3) \citet{rn02}, 4) Upper limit to orbit, 5) \citet{soh07}, 6) \citet{hbp01}, 7) \citet{hob08}, 8) \citet{rsa98}. }
\label{allastrometry}
\end{deluxetable*}

Characterization of the binary nature of A-star systems becomes doubly interesting when the host stars are also rapidly rotating. Recent interferometric imaging of several rapidly rotating A-stars have revealed imaging of their surfaces \citep{mzp07,zmp09}.  $\alpha$ Oph, a prototypical rapid rotator, is rotating at $\sim$89\% of its breakup velocity.  Interferometric surface imaging of this star clearly shows its pronounced oblateness: the star's equatorial radius is 20\% larger than its polar radius, corresponding to a $\sim$1840K temperature differential \citep{zmp09}.  This 2-3 M$_\odot$ star has also been the target of extensive asteroseismic monitoring carried out by the MOST satellite, discovering $\sim$50 pulsational modes (J. D. Monnier, private communication).

Derivation of the component masses for this system is a critical check against investigations of the host star's rapid rotation and asteroseismology. 
Discrepancies between the dynamically-derived mass of the host star and that allowed by models of rapid rotators, may illuminate potential complications with rotator models, e.g. those that do not take into account differential rotation of the star.  
For all these reasons, deriving a more accurate orbit is crucial for future studies to reconcile the orbital dynamics of the companion star with the physics dictating the rotation of the host star.  
In this paper, we assemble data from a number of high-contrast imaging programs over the past eight years.  We take advantage of these recent measurements along with archival data to further constrain the orbit of the companion, as well as the system mass.

\section{Astrometric Observations} 
 Table~\ref{allastrometry} presents astrometric measurements of $\alpha$ Oph obtained using several instruments.  Epochs one through five are previously measured one-dimensional speckle data from \citet{klm89}.  For the sixth epoch of observations, we adopt the radial separation (770$\pm$ 40 mas) and position angle (243.7$\pm$ 3.0$^\circ$) measurements reported in \citet{bmm01}. 
\citet{bmm01} mistakenly reported the position angle in degrees measured West of North.  \citet{g05} noticed the problem, but mistakenly assumed it was a quadrant problem and changed the value by 180$^\circ$.  We use the correct measurement position angle of 243$^\circ$ (A. Boccaletti, private communication).

\begin{figure}[ht]
\center
\resizebox{.96\hsize}{!}{\includegraphics[angle=-90]{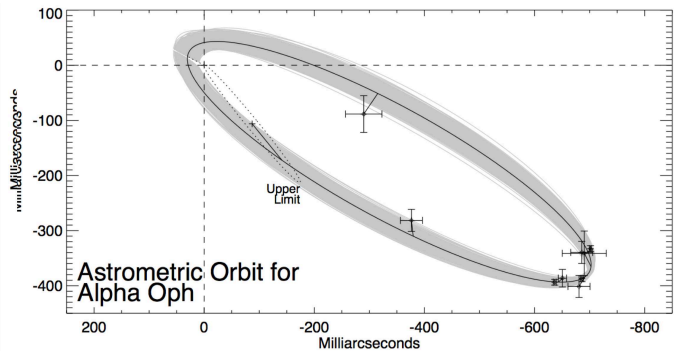}}
  \caption{The visual orbit of the $\alpha$ Oph system.  The solid line is the best fit solution while the grey band shows the allowed orbits based on Monte Carlo sampling of the parameter uncertainties.  The primary is located at (0,0) and the astrometric data from Table~\ref{allastrometry} are plotted with the positional uncertainties.  East is left, and North is up.}
\label{orbit} 
\end{figure}

 Epochs seven and eight are comprised of visible AO measurements obtained from the 3.63m AEOS telescope on Haleakala, Hawaii \citep{rn02,rtb05}. Epochs nine and eleven were obtained using ``The Lyot Project'' \citep{soh07,hos07,lsh10}, a diffraction-limited classical Lyot coronagraph \citep{l39,skm01} working in the infrared and recently decommissioned at AEOS.  

\begin{figure*}[ht]
\center
\resizebox{1.03\hsize}{!}{\includegraphics{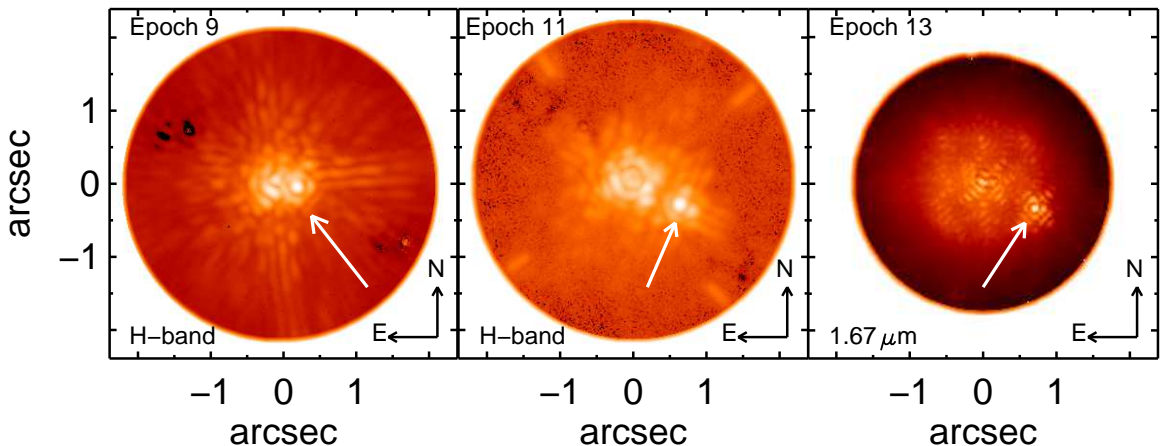}}
  \caption{Examples of data taken at the three epochs listed at the upper left hand corner of each panel.  The left and middle images were taken with a near-IR camera with a coronagraph at the AEOS telescope.  The right hand panel is a single slice extracted from a data cube from the high contrast integral field spectrograph at Palomar Observatory \citep{hob08}.  The companion, $\alpha$ Oph B, is marked with a white arrow in all three epochs, and is shown near apastron in the third panel.}
\label{three_epochs} 
\end{figure*}

The thirteenth epoch of observations of the $\alpha$ Oph system was obtained using a recently commissioned coronagraph integrated with an integral field spectrograph (IFS) spanning the $J$ and $H$-bands (1.10$\mu$m - 1.75$\mu$m) on the 200-in Hale Telescope at Palomar Observatory \citep{hob08}. This instrument package, called ``Project 1640'' is mounted on the Palomar AO system \citep{dbp98,tdb00}, and is a dedicated high contrast imaging instrument providing low resolution spectra  ($\lambda/\Delta\lambda$)$\sim$30---60 with a lenslet-based IFS and  Apodized-Pupil Lyot coronagraph \citep{s05}, an improvement of the classical Lyot coronagraph \citep{skm01}.  Epoch 14 observations were made at the Canada-France-Hawaii Telescope \citep{rsa98}. The observations comprising epochs 10, 12, and 15 were obtained with the Palomar AO system and the PHARO infrared camera \citep{hbp01}.  Astrometric points from epochs six through fifteen are shown in Figure~\ref{orbit}, and representative examples of data from epochs 9, 11, and 13 are shown in Figure~\ref{three_epochs}.  


 \begin{deluxetable}{ll}
\tabletypesize{\scriptsize}
\tablecaption{Binary Orbit and Derived Properties}
\tablewidth{0pt}
\tablehead{\colhead{Parameter}  & \colhead{Value}}
\startdata                                     
{\bf From Visual Orbit Only:} &    \\
  \hspace{0.05in}Semi-major axis $a$ (mas)	        &	427$^{+20}_{-13}$ \\
  \hspace{0.05in}Eccentricity                                         &       0.92 $\pm$ 0.03 \\
  \hspace{0.05in}Inclination (degs)                              &       125$^{+6}_{-9}$ \\
  \hspace{0.05in}$\omega$ (degs)                              &       162 $\pm$14 \\
  \hspace{0.05in}$\Omega$ (degs)                             &       232$\pm$9 \\
   \hspace{0.05in}Period (days)                                  &       3148.4 (fixed) \\
  \hspace{0.05in}Epoch of periastron (JD)                &    2452888 $\pm$ 53\\
                                                               &                            \\
{\bf From Astrometric Photocenter}  &                \\  
 {\bf Motion (Using Visual Orbit Above):}    & \\
  \hspace{0.05in}Mass Ratio (Primary/Secondary):    &    2.76 $^{+.43}_{-.27}$ \\
  \hspace{0.05in}Parallax (mas)                                     & 69.1 (fixed)   \\
  \hspace{0.05in}Fixed proper motion: E,N (mas yr$^{-1}$)  &  123.3, -227.0  \\
   \hspace{0.05in}System Mass (M$_\odot$):            & $3.25^{+.35}_{-.40}$ \\
   \hspace{0.05in}$\alpha$ Oph A (M$_\odot$):       & $2.40^{+.23}_{-.37}$ \\
  \hspace{0.05in}$\alpha$ Oph B (M$_\odot$):      & 0.85$^{+.06}_{-.04}$  \\
 \enddata
\label{binaryorbit}
\end{deluxetable}

\section{Orbital Characterization}\label{orbitchar}
We fitted an orbit to the astrometric data in Table~\ref{allastrometry} including the speckle data from \citet{klm89} \footnote{We needed to flip the signs on x,y from this work to be consistent with other astrometry measurements.}.  We used both a Levenberg-Marquardt method (using IDL procedure MPFIT developed by C Markwardt \footnote{http://cow.physics.wisc.edu/$\sim$craigm/idl/idl.html}), and also the IDL routine AMOEBA \citep{ptv92},  arriving at the same global solution.  Comparing our predicted astrometry with the data, we find a reduced $\chi^2$ of 0.7.  The best-fit orbital elements can be found in Table~\ref{binaryorbit}, where we have fixed the quadrant of $\omega$ using the radial velocity data of \citet{klm89}. In order to calculate errors for the orbital elements, we carried out a Monte Carlo simulation, allowing all parameters to vary in the fitting except for the period, which we fixed to 8.62~years.  The results of the error analysis can be found in Table~\ref{binaryorbit}.

\begin{deluxetable}{lcccc}
\tabletypesize{\scriptsize}
\tablecaption{Apparent Magnitudes for the $\alpha$ Oph system}
\tablewidth{0pt}
\tablehead{\colhead{Member}  & \colhead{$I$} & \colhead{$J$} & \colhead{$H$} & \colhead{$K$}}
\startdata                                     
        $\alpha$ Oph A\footnotemark[1] & $1.90               $ & $1.83\pm0.28$ & $1.72\pm0.18$ & $1.68\pm0.21$ \\
        $\alpha$ Oph B\footnotemark[2] & $6.72\pm0.10$ & $6.02\pm0.310$ & $5.44\pm0.206$ & $5.25\pm0.236$ 
 \enddata
 \footnotetext[1]{$I$-band photometry: \citet{mlc03}, $JHK$: \citet{csv03}}
 \footnotetext[2]{$I$-band photometry was obtained using the AEOS VisIm instrument and $JHK$ photometry with the PHARO infrared camera at Palomar.}
\label{phot}
\end{deluxetable}

\begin{figure}[ht]
\center
\resizebox{1.0\hsize}{!}{\includegraphics[angle=-90]{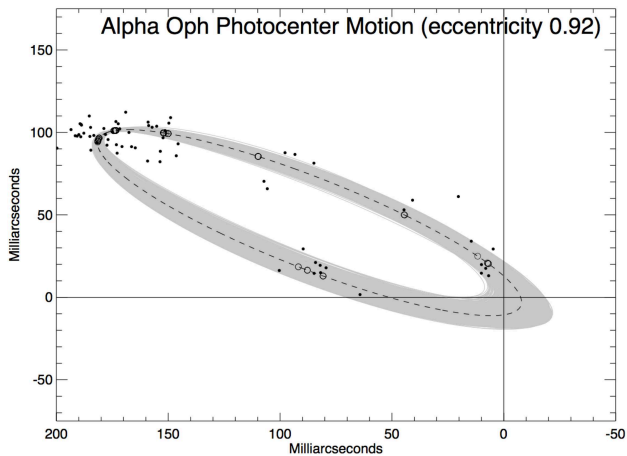}}
  \caption{The relative photocenter positions of $\alpha$ Oph from MAP observations \citep{g87} reported by \citet{g05} after subtracting out the system proper motion and an estimate of the center of mass.  With the orbit fit shown in Figure~\ref{orbit}, the component mass ratio is obtained by fitting the photocenter motion shown here.  }
\label{photocenter} 
\end{figure}

Using the full covariant set of parameters from the monte carlo study, we were then able to determine the component mass ratio by fitting  the photocenter motion measured with the MAP instrument \citep{g05}.   In order to carry out a fit to the absolute position of the photocenter, we needed to constrain a few important parameters using independent sources. Firstly, we adopted a proper motion of 123.3 mas yr$^{-1}$ (East) and -227.0 mas yr$^{-1}$ (North) from the FK5 catalog.  Note that other proper motion catalogs with shorter time baselines contain significantly different values for $\alpha$ Oph, presumably having been corrupted by orbital motion of the primary.  We also adopted a parallax of 69.1 mas from \citet{mm98}.  Note that the errors in these three quantities have been neglected in the following analysis (errors from our visual orbit dominate uncertainties in the mass ratio).  Figure~\ref{photocenter} shows these relative photocenter offsets after subtracting out the system proper motion and an estimate of the center of mass. Lastly, we had to assume a flux ratio at the effective wavelength of the MAP experiment \citet{g87}, similar to an $R$-band filter.  Based on multi-wavelength detections of the companion at $I$, $J$, $H$, and $K$-bands, we have estimated an $R$-band flux ratio of 5.3 magnitudes (factor of 130) assuming the companion is a mid/late K star (See Section~\ref{sed}).  Using the results from the two analyses, we constrain the masses to be $2.40^{+0.23}_{-0.37}$ M$_\odot$ and $0.85^{+0.06}_{-0.04}$ M$_\odot$ for $\alpha$ Oph A and $\alpha$ Oph B, respectively.  Table~\ref{binaryorbit} contains the results of this orbital fitting, along with the full set of the fit parameters.

\section{Analysis}                         
\subsection{Companion Spectral Energy Distribution}\label{sed}
\begin{figure}[ht]
\center
\resizebox{1.08\hsize}{!}{\includegraphics{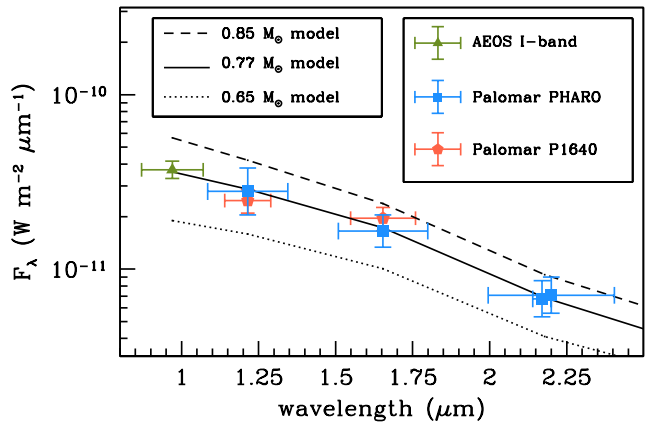}}
  \caption{Broadband photometry for $\alpha$ Oph B from three of the observing programs described in this paper.  Also shown are three theoretical broadband $I$, $J$, $H$, and $K$-band flux curves based on the mass-luminosity models of \citet{sdf00}. Each curve is an interpolation between these values.  The best fitting model is shown (solid line--0.77 M$_\odot$), as is the dynamically determined mass of 0.85 M$_\odot$, and a 0.65 M$_\odot$ curve for reference. The horizontal error bars reflect the width of each bandpass.}
\label{spectrum} 
\end{figure}

Table~\ref{phot} presents the broadband apparent magnitudes for $\alpha$ Oph B obtained with the PHARO infrared camera at Palomar, as well as the AEOS VisIm camera. The spectral energy distribution for $\alpha$ Oph B is shown in Figure~\ref{spectrum}, accounting for the system's 0.834$\pm$0.024 distance modulus \citep{g05}. The broadband magnitudes listed in Table~\ref{phot} are the result of a simultaneous photometric fit to each binary component which returns position angle, relative separation as well as relative brightnesses. Figure~\ref{spectrum} shows $I$, $J$, $H$, Br$_\gamma$, and $K$-band photometry. Also shown are broadband $J$ and $H$-band photometric values obtained with the Project 1640 IFS at Palomar Observatory. The Project 1640 observations were calibrated using reference star spectra obtained from the IRTF spectral library \citep{crv05,rcv09} in the same manner as described in \citet{hob10}.  However, these observations were  gathered during the first light observations with this new instrument, and hence the calibration dataset was obtained before a full understanding of an effective calibration process was mature.  


Also shown in Figure~\ref{spectrum} are theoretical broadband $I$, $J$, $H$, and $K$-band fluxes based on the models of \citet{sdf00}. The curves in Figure~\ref{spectrum} are interpolations between these values.  At a spectral resolution of 30-60, late-G to early-M star spectra are very similar, preventing robust discrimination between them.  Rather than plotting template spectra over the data for comparison, we have instead chosen to use these model-based interpolated mass-luminosity curves to allow discrimination of the best fit mass value for $\alpha$ Oph B.    The best fit curve to the measured fluxes corresponds to a mass of 0.77 M$_\odot$.  Taking into account our photometric errors as well as the 0.024 magnitude distance modulus uncertainty quoted by \citet{g05}, we deduce an uncertainty of $\pm$0.05 M$_\odot$ in our best fit value.  This value of 0.77$\pm$0.05 M$_\odot$ is very similar to the 0.78$\pm$0.058 M$_\odot$ reported by \citet{g05}, and within 1.3$\sigma$ of our 0.85$^{+.06}_{-.04}$ M$_\odot$ dynamical mass discussed in Section~\ref{orbitchar}.  
Moreover, in this mass regime ($\sim$0.7---0.8M$_\odot$), \citet{hw04} have demonstrated that mass-luminosity models, including those produced by \citet{sdf00}, are only inaccurate at the few percent level. 
Due to the aforementioned uncertainties in the calibration of the Palomar Project 1640 photometry, only the Palomar PHARO data and the AEOS $I$-band data were used in the fit. Figure~\ref{spectrum} also shows a curve corresponding to the dynamical best-fit mass value of 0.85 M$_\odot$, as well as a curve corresponding to 0.65 M$_\odot$ for reference.  
Comparing our derived near-infrared $(J-H)$ and $(H-K)$ colors with those documented in \citet{bb88} puts the spectral type of the companion between a K5V and a K7V. 

\subsection{Spin-Orbit Mutual Inclinations}
Our measurement of the orbital plane of the Alpha Oph binary system can be compared to the angular momentum of the rapidly-rotating primary star. This is analogous in practice to measuring the relative orbital inclinations of triple or quadruple system \citep[e.g.,][]{mlf08} which can reveal hints to the formation and/or evolution of system dynamics \citep[e.g.,][]{stt02}.   In order to carry out this comparison, we convert the spin parameters from \citet{zmp09} into an equivalent $\Omega=216.12\pm1.23\arcdeg$, $i=92.30\pm0.43\arcdeg$ that would characterize a circular orbit in the star's equatorial plane. Note these values are affected by orbital degeneracies since we have not incorporated any surface or orbital velocities into the estimate. However, because the primary spin geometry is nearly edge-on, these degeneracies do not matter much.We find that the mutual angle of inclination is $\Phi=36\pm6\arcdeg$ (or $\Phi=144\pm6\arcdeg$ for the opposite spin polarity).  Interestingly, this lies close to the predicted peak inclination (40\arcdeg) that arises from  Kozai cycles in the context of an inner short-period binary \citep{ft07}, echoing findings of  \citet{mlf08}.  However, the primary $\alpha$~Oph~A has no known inner companion and so this mechanism does not seem to be applicable. 

\section{Discussion and Conclusions}
We have obtained an orbital fit for the $\alpha$ Oph system and derived component masses of $2.40^{+.23}_{-.37}$  and 0.85$^{+.06}_{-.04}$ M$_\odot$ for $\alpha$ Oph A, and $\alpha$ Oph B, respectively.  Our estimation of the mass of $\alpha$ Oph A is lower than the 2.84$\pm$0.19 M$_\odot$ value quoted by \citet{g05}.  However, our Monte Carlo approach yields more conservative estimates of the uncertainties, overlapping with other works.  Specifically, the 2.1 M$_\odot$ value for $\alpha$ Oph A quoted by \citet{zmp09} and measured using an HR diagram overlaid with isochrones, is allowed at this work's 1$\sigma$ level, while it is only allowed at the 3.5$\sigma$ level in the \citet{g05} study.  Our work demonstrates that, especially with a high eccentricity orbit, careful estimates of stellar parameters are essential.  

We anticipate that this dynamical estimation of the mass of $\alpha$ Oph A will aid forthcoming studies focused not only on modelling the rapid rotation of $\alpha$ Oph A, but also of the host star's asteroseismological properties (J. D. Monnier, private communication). The mass value derived here may help to distinguish between models that assume solid-body rotation of the star, or have some degree of differential rotation beneath the photosphere. 

Moreover, we have compiled the most complete set of photometric data yet for $\alpha$ Oph B from several observing programs to obtain an estimate of the companion mass through the use of theoretical models. The 0.77$\pm$0.05 M$_\odot$ estimation of the companion mass derived photometrically is marginally consistent with the 0.85$^{+.06}_{-.04}$ M$_\odot$ value for $\alpha$ Oph B.  Our infrared colors and the corresponding best fit mass-luminosity curve are consistent with a mid/late-K dwarf classification for the companion. 

Finally, based on our fit, $\alpha$ Oph B will be passing through periastron on 2012 April 19, with an uncertainty on this date of $\pm$ 53 days.  However, the separation between the two objects should be $\sim$50 mas from 2012 March 24 to 2012 May 21. Such a separation will be ideal for existing state-of-the-art interferometers such as the CHARA Array or the Very Large Telescope Interferometer.  Intensive astrometric monitoring of this system and new radial velocity observations, especially during periastron, can help to constrain the mass of $\alpha$ Oph A to within a few percent.

\acknowledgments 
This work was performed in part under contract with the California Institute of Technology (Caltech) funded by NASA through the Sagan Fellowship Program.The Lyot Project is based upon work supported by the National Science  Foundation under Grant Nos. AST-0804417, 0334916, 0215793, and 0520822.  The Lyot Project grateful acknowledges the support of the  US Air Force and NSF in creating the special Advanced Technologies  and Instrumentation opportunity that provides access to the AEOS  telescope.  The Lyot Project is also grateful to the Cordelia Corporation, Hilary  and Ethel Lipsitz, the Vincent Astor Fund, Judy Vale and an anonymous  donor who initiated the project.   A portion of the research in this paper was carried out at the Jet Propulsion Laboratory, California Institute of Technology, under a contract with the National Aeronautics and Space Administration.   Thanks also to Anthony Boccaletti for clarification of his epoch of the $\alpha$ Oph astrometry. We thank Willie Torres for help checking the notation of our orbital elements, and also to the anonymous referee for several helpful comments. 


\bibliography{/scr/shinkley/Workspace/papers/MasterBiblio_Sasha}
\bibliographystyle{/scr/shinkley/Workspace/mypapers/IFU_PASP_paper/apj.bst}

\end{document}